\documentclass{article}
\usepackage{spconf,amsmath,amssymb,graphicx}
\usepackage{multirow}
\usepackage{booktabs, bbding}
\usepackage{cite}

\title{Improving Transformer-based End-to-End Speaker Diarization by Assigning Auxiliary Losses to Attention Heads}

\name{Ye-Rin Jeoung, Joon-Young Yang, Jeong-Hwan Choi, and Joon-Hyuk Chang}
\address{
  Department of Electronic Engineering\\
  Hanyang University, Seoul, Republic of Korea}

\begin{document}
\ninept
\maketitle
\begin{abstract}
Transformer-based end-to-end neural speaker diarization (EEND) models utilize the multi-head self-attention (SA) mechanism to enable accurate speaker label prediction in overlapped speech regions. In this study, to enhance the training effectiveness of SA-EEND models, we propose the use of auxiliary losses for the SA heads of the transformer layers. Specifically, we assume that the attention weight matrices of an SA layer are redundant if their patterns are similar to those of the identity matrix. We then explicitly constrain such matrices to exhibit specific speaker activity patterns relevant to voice activity detection or overlapped speech detection tasks. Consequently, we expect the proposed auxiliary losses to guide the transformer layers to exhibit more diverse patterns in the attention weights, thereby reducing the assumed redundancies in the SA heads. The effectiveness of the proposed method is demonstrated using the simulated and CALLHOME datasets for two-speaker diarization tasks, reducing the diarization error rate of the conventional SA-EEND model by 32.58$\%$ and 17.11$\%$, respectively.
\end{abstract}
\begin{keywords}
speaker diarization, end-to-end neural diarization, multi-head attention, auxiliary loss
\end{keywords}
\section{Introduction}
\label{sec:intro}
Speaker diarization, which aims at determining ``who spoke when", is useful for analyzing multi-speaker long-form audio obtained from telephone conversations \cite{kenny2010diarization}, interviews, official meetings \cite{janin2003icsi,renals2008interpretation}, and TV programs \cite{vallet2012multimodal}.
Conventionally, the speaker diarization technique is implemented by combining several independently designed modules, such as voice activity detection (VAD), overlapped speech detection (OSD), speaker embedding extraction, and clustering.
However, this approach has limitations in that the constituent modules are not jointly optimized for the speaker diarization task, but rather for the relevant subtasks.
Moreover, overlapped speech regions are particularly difficult to be dealt with because the speaker embedding extractor is typically trained to embed single-speaker information.

To overcome these limitations, end-to-end neural speaker diarization (EEND) approaches have been proposed, which directly predict speaker activity labels from a sequence of acoustic features in an end-to-end manner.
Specifically, the self-attentive (SA) EEND model \cite{fujita2019end2} comprises a stack of transformer \cite{vaswani2017attention} layers and produces frame-level speaker posteriors as outputs.
The SA-EEND model was trained using only the binary cross-entropy (BCE) loss between the speaker labels and predictions obtained from the last layer using a permutation invariant training (PIT) method \cite{fujita2019end2}.
However, the SA-EEND model thus trained leads to redundancies in the lower layers (i.\,e., the contributions of the lower layers to the target task are not significant), particularly for the increased number of transformer layers \cite{yu2022auxiliary}.
To address this problem, \cite{yu2022auxiliary} proposed adding residual connections between consecutive transformer layers and applying auxiliary BCE losses to intermediate transformer layers.
This modified architecture was denoted as RX-EEND \cite{yu2022auxiliary} and substantially improved the performance compared with the original SA-EEND model.

In this study, we aim to improve the performance of the SA-EEND model from a perspective different from that described in \cite{yu2022auxiliary}.
Specifically, we focus on the observation presented in \cite{fujita2019end2} that the attention weight matrices of the SA-EEND model tend to exhibit patterns similar to an identity matrix, which we assume to be redundant for the speaker diarization task.
Accordingly, inspired by a recent study \cite{lee22b_interspeech}, we propose explicitly constrained attention weight matrices that exhibit specific patterns derived from auxiliary tasks relevant to speaker diarization.
In \cite{lee22b_interspeech}, to promote an SA head to accurately attend to short speech utterance within long-duration non-speech regions, an SA head was constrained to exhibit speech activity patterns derived from true VAD labels.
Rather, in this study, we focus on VAD and OSD in a multi-speaker situation, both of which are important for achieving good speaker diarization performance \cite{bredin2020pyannote,takashima2021end,bredin2021end}.
Patterns derived from the true VAD or OSD labels of overlapped speech utterances are used to constrain the attention weight matrices of the SA-EEND model in the form of auxiliary loss functions.
Consequently, we expect the attention weight matrices to exhibit target-task-relevant patterns rather than identity-like matrices that assign equal weights to the entire input sequence.
The effectiveness of the proposed approach is verified using both simulated and real datasets for a two-speaker diarization task.

\section{Review of SA-EEND}
In this section, we briefly review the SA-EEND \cite{fujita2019end2} model as a baseline system.
The SA-EEND model predicts the speaker label sequence $Y = [\mathbf{y}_1,\cdots,\mathbf{y}_T]$ given a $T$-length feature vector sequence $X=[\mathbf{x}_1,\cdots,\mathbf{x}_T]$.
Speaker label $\mathbf{y}_t = [y{_{t,1}},\cdots,y{_{t,S}}]^{\top} \in \{0,1\}^S$ denotes the activities of the $S$ speakers at time frame index $t$.
After passing through a linear layer and layer normalization (LN) \cite{ba2016layer}, $\mathbf{x}_t$ is converted to an embedding $\mathbf{e}_t^0$, and subsequently introduced to the stacked transformer encoder blocks.
\begin{align}
    \mathbf{e}_t^0 &= \textup{LN}(\textup{Linear}(\mathbf{x}_t))\ ,\ \mathbf{e}_t^0\in \mathbb{R}^D, \\
    E^p &= \textup{EncoderBlock}_p^D(E^{p-1}) , 1\le p\le P,
\end{align}
where $\textup{Linear}(\cdot)$ denotes a linear layer, $p$ indexes the encoder blocks, and $E^p=[\mathbf{e}_1^p, \cdots ,\mathbf{e}_T^p]$ denotes the embedding sequence.
Each transformer encoder block consists of multi-head self-attention (MHSA) and fully connected feed-forward networks.
SA uses a scaled dot-product \cite{vaswani2017attention} to produce frame-level attention weights that consider global feature relations.
\begin{equation}
    Attention(Q,K,V) = \textup{softmax}({QK^{\top} \! \big/ \sqrt{d}})V = AV, \label{eqn:attention}
\end{equation} 
where $Q$, $K$, and $V$ $\in \mathbb{R}^{T\times d}$ denote the query, key, and value, respectively \cite{vaswani2017attention}; $d$ is the dimension of the hidden space; and $A \in \mathbb{R}^{T\times T}$ denotes the attention weight matrix.
The output of the SA layer is subsequently passed to a feed-forward network consisting of two linear transforms with the ReLU activation \cite{agarap2018deep}.
Finally, after passing through all encoder blocks, 
the speaker posteriors for $S$ speakers, $\hat{\mathbf{y}}_t = [\hat{y}{_{t,1}},\cdots,\hat{y}{_{t,S}}]$, at time frame $t$ are obtained as follows:
\begin{equation}
    \hat{y_t} = \textup{sigmoid}(\textup{Linear}(\textup{LN}(\mathbf{e}_t^P))).
\end{equation} 

To train the SA-EEND model, the loss function, $\mathcal{L}_d$, between the prediction $\hat{\mathbf{y}}_t$ and ground truth label $\mathbf{y}_t$ is calculated as follows:
\begin{align}
    H({y}_{t,s},\hat{y}_{t,s}) &= -y_{t,s}\ln\hat{y}_{t,s}-(1-y_{t,s})\ln(1-\hat{y}_{t,s}), \\
    \mathcal{L}_d &= {1 \over TS}\, {\min_{\phi_{1},\cdots,\phi_{S} \in \Phi_{S}}} \displaystyle\sum_{t=1}^T \displaystyle\sum_{s=1}^S {H({y}_{t,s}^{\phi_{s}},\hat{y}_{t,s})},
\end{align}
where $H(\cdot,\cdot)$ denotes the BCE loss, $\Phi_{S}$ represents all possible permutations of the speakers, and $\mathbf{y}_{\phi_{s}}^{}=[{y}_{1,s}^{\phi_{s}},\cdots,{y}_{T,s}^{\phi_{s}}]\in \{0,1\}^T$ is the speaker label sequence following the speaker permutation $\phi_{s}$.

\section{Proposed auxiliary losses} \label{sec:proposed}

\subsection{Speaker-wise VAD loss} \label{sec:proposed:svad_loss}
To help SA heads learn the voice activity patterns of different speakers, we define a new auxiliary loss called speaker-wise VAD (SVAD) loss.
To compute the SVAD loss, the speaker label sequence, $\mathbf{y}_{\phi_{s}}^{}$, is used to generate the target masks, $\mathcal{M}_{s} = {\mathbf{y}_{\phi_{s}}^{\top}}{\mathbf{y}_{\phi_{s}}^{}} (1 \leq s \leq S)$, which are used to constrain the attention weight matrices as follows:  
\begin{align}
 \mathcal{L}_{S} &= \displaystyle\sum_{s=1}^{S} \mathcal{L}(\mathcal{M}_{s},A_{s}^{h}) = \displaystyle\sum_{s=1}^{S} {1\over T^2}\displaystyle\sum_{i=1}^{T}\displaystyle\sum_{j=1}^{T} H(m_{ij},a_{ij}), \label{eqn:svad_loss}
\end{align}
where the attention weight matrix $A_{s}^{h}$, selected for $\mathcal{M}_{s}$, is calculated from the $h$-th SA head, and $m_{ij}$ and $a_{ij}$ denote the ($i$,$j$)-th entries of $\mathcal{M}_{s}$ and $A_{s}^{h}$, respectively. 
Because the objective of speaker diarization is to produce the VAD label for each speaker, we expect the patterns of the target masks, $\mathcal{M}_{s}$, to help the SA heads learn effective attention scheme that does not lead to identity-like patterns.

\begin{figure}[t]
  \centering
  \includegraphics[width=\linewidth]{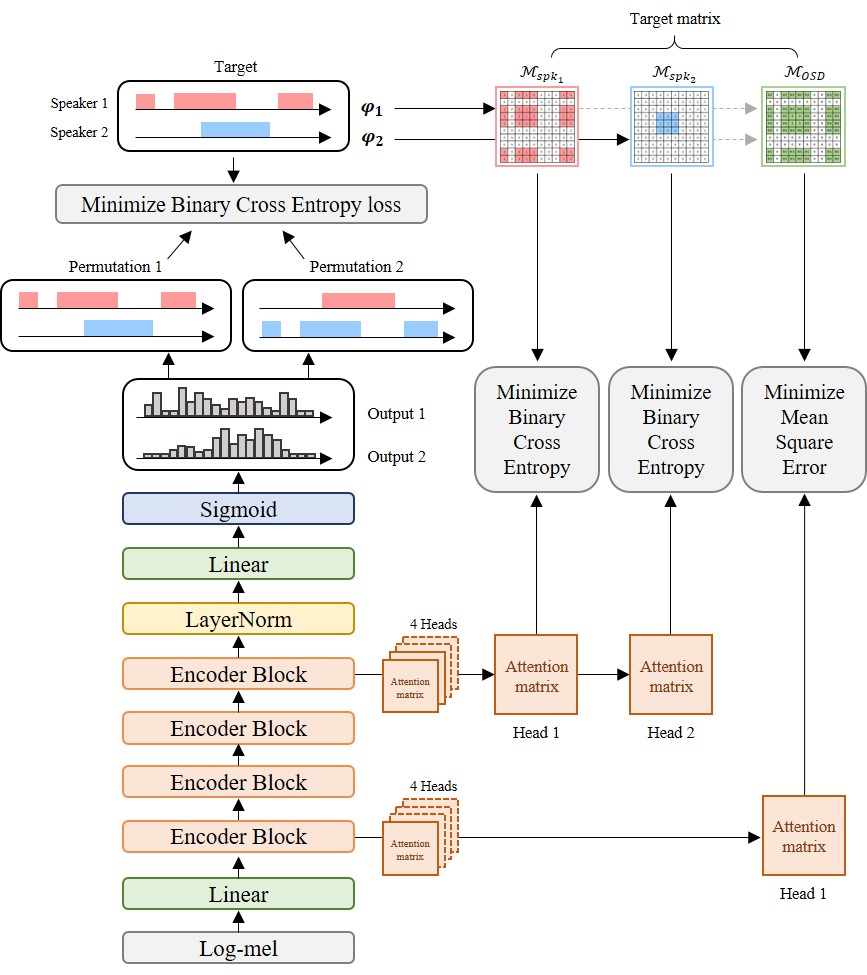}
  \caption{SA-EEND model with proposed auxiliary losses for two speakers}
  \label{fig:speech_production}
\end{figure}

\subsection{OSD loss} \label{sec:proposed:osd_loss}
To help SA heads learn the overall speech activity patterns, while differentiating overlapped speech and single-speaker regions, we define an auxiliary loss called OSD loss.
The OSD label sequence, $\boldsymbol{{\psi}}=[{\psi}_1,\cdots,{\psi}_T]$, is first defined as follows:
\begin{equation}
{\psi}_t =\begin{cases}
0 &\text{if}\ \textrm{$t$ is non-speech frame} \\
k &\text{if}\ \textrm{$t$ is single-speaker speech frame} \\
1 &\text{if}\ \textrm{$t$ is overlapped speech frame} \\ 
\end{cases}
. \label{eqn:osd_loss}
\end{equation} 
Then, similar to that described in Section \ref{sec:proposed:svad_loss}, the target masks, $\mathcal{M}_{OSD}= \boldsymbol{{\psi}}^{\top}\boldsymbol{{\psi}}$, are used to define the OSD loss.
\begin{equation}
\mathcal{L}_{O}={1\over T^2}\displaystyle\sum_{i=1}^{T}\displaystyle\sum_{j=1}^{T}(n_{ij}-a_{ij})^2,
\end{equation}
where $n_{ij}$ denotes the ($i$,$j$)-th entry of $\mathcal{M}_{OSD}$, and $a_{ij}$ is the corresponding entry of the attention weight matrix to which the OSD loss is applied.
Through this, we expect the attention weights to implicitly differentiate the single-speaker and overlapped speech regions, instead of having equal weights that leads to identity-like patterns.

\subsection{Model training with proposed auxiliary losses} \label{sec:proposed:training}
Because we assume the identity matrix to be redundant, we select the SA head, whose attention weight matrix is most similar to the identity matrix pattern, for applying the proposed auxiliary losses.
To implement this method, we calculate the traces of the attention weight matrices for each head and arranged those values in a descending order to select the redundant SA heads.

The diarization loss, $\mathcal{L}_d$, and auxiliary losses, $\mathcal{L}_{S}$ and $\mathcal{L}_{O}$, are assigned together to help each head distinguish not only the presence of a speech but also the speech of each speaker.  
Finally, the total loss function for training our proposed system is defined as 
\begin{equation}
\mathcal{L}_{Total} = \mathcal{L}_d + \alpha \mathcal{L}_{S} + \beta \mathcal{L}_{O}, \label{eqn:L_total}
\end{equation}
where $\alpha$ and $\beta$ are the hyperparameters that indicate the degree of each auxiliary loss to be applied. 

\section{Experimental setup}

\subsection{Datasets}
The simulated dataset, Sim2spk, was created according to the speech mixture simulation algorithm described in \cite{fujita2019end2}, with an overlap ratio of 34.4\%.
Specifically, Sim2spk was generated based on the Switchboard-2 (Phases I, II, and III), Switchboard Cellular (Part1 and Part2) \cite{godfrey1992switchboard}, and NIST Speaker Recognition Evaluation (2004, 2005, 2006, and 2008) corpora \cite{alvin2004nist, martin2005nist, przybocki2006nist, martin2009nist}, all of which were sampled at 8 kHz. 
All utterances in Sim2spk were composed of two distinct speakers, each of which was generated by randomly selecting 10 to 20 utterances from each speaker.
Following \cite{fujita2019end2}, each utterance was convolved with a randomly selected simulated room impulse response with a probability of 0.5 and was added with background noise samples taken from the MUSAN dataset \cite{snyder2015musan}.
For the evaluation dataset, both simulated and real datasets were prepared.
The simulated datasets were created using the same method as described above, but with overlap ratios of 34.4\%, 27.3\%, and 19.6\%.
For the real evaluation dataset, a set of two-speaker telephone conversation utterances in the CALLHOME (CH) \cite{LDC2001S97} dataset was used.

\begin{table}[t]
\caption{Statistics of two-speaker datasets}
  \label{tab:example}
  \centering
\resizebox{\columnwidth}{!}{%
\begin{tabular}{ccccc}
\toprule[1.2pt]
\multirow{2.5}{*}{\textbf{Datasets}} & \textbf{Train}   & \textbf{Adapt} & \multicolumn{2}{c}{\textbf{Evaluation}} \\ \cmidrule(lr){2-2} \cmidrule(lr){3-3} \cmidrule(lr){4-5} \\[-1em]
& Sim2spk & CH    & Sim2spk          & CH   \\[-1em] \\ \hline \\[-1em] \\[-1em]
$\#$ mixtures                & 100,000 & 155   & 500 / 500 / 500      & 148   \\ \\[-1em]
overlap ratio (\%)         & 34.4    & 14.0  & 34.4 / 27.3 / 19.6   & 13.1  \\ 
\bottomrule[1.2pt]
\end{tabular}
}
\end{table}

\subsection{Training and evaluation}
The SA-EEND model \cite{fujita2019end2} with four transformer encoder blocks was used as the baseline system, where each block used four heads for MHSA.
The input features were 23-dimensional log-scaled mel-filterbank energies extracted with a 25 ms frame length and 10 ms frame shift \cite{fujita2019end2}.
A single frame of feature was spliced with 7 left and 7 right frames, and the sequence of features extracted from an utterance was temporally downsampled by a factor of $10$ \cite{fujita2019end2}.
%
The proposed auxiliary losses were applied to the SA-EEND model as described in Section \ref{sec:proposed}.
Instead of determining the optimal values of $\alpha$ and $\beta$ in Eq.\,(\ref{eqn:L_total}), $\alpha$ was set to $1$ when $\mathcal{L}_{S}$ was applied, and $0$ otherwise; $\beta$ was set in the same manner.
For the OSD loss, we set $k=\sqrt{0.5}$ in Eq.\,(\ref{eqn:osd_loss}) so that the corresponding entry of $\mathcal{M}_{OSD}$ becomes $0.5$.
The EEND model was pretrained for 200 epochs using Sim2spk dataset, and subsequently adapted for 200 epochs using the CH two-speaker dataset.
The Adam optimizer \cite{kingma2014adam} was used with 100,000 warmup steps \cite{vaswani2017attention} in the pretraining stage, and the learning rate was set to $0.00001$ during adaptation.
The final speaker activity predictions were obtained using a threshold of $0.5$ and median filtering with a window size of 11 frames.
Diarization error rate (DER) \cite{fiscus2007rich} was used as the evaluation measure, and 0.25s of collar tolerance was used at the beginning and end of each segment.

\renewcommand{\arraystretch}{0.95}
\begin{table}[t]
\caption{Performance comparison of different EEND approaches in terms of DER ($\%$). $\dagger$: Our implementation}
  \label{tab:der_comparison}
  \centering
  \begin{tabular}{ccccccc}
  \toprule[1.2pt]
  \multicolumn{2}{c}{\multirow{2.5}{*}{\textbf{Method}}}                                                                                        & \multicolumn{3}{c}{\textbf{Sim2spk}} & \textbf{Real} \\ \cmidrule(lr){3-5} \cmidrule(lr){6-6}
  \multicolumn{1}{c}{}  & & $\rho$ = 34.4       & $\rho$ = 27.3       & $\rho$ = 19.6       & CH            \\ \\[-1em] \hline
  \\[-1em]
  \\[-1em]
  \multicolumn{2}{c}{SA-EEND \cite{fujita2019end2}}                                                                                                                & 7.91       & 8.51       & 9.51       & 13.66         \\ \\[-1em]
  \multicolumn{2}{c}{SA-EEND$^{\dagger}$}                                                                                                                & 6.30       & 6.14       & 6.17       & 10.46         \\ \\[-1em]
  \multicolumn{2}{c}{RX-EEND \cite{yu2022auxiliary}}                                                                                                                & 4.18       & 3.93       & 4.01       & 9.17         \\ \\[-1em]
  \multicolumn{2}{c}{RX-EEND$^{\dagger}$}                                                                                                                & 4.28       & 3.76       & 3.86       & 9.04         \\ \\[-1em]
  \hline
  \\[-1em]
  \\[-1em]
  \multicolumn{2}{c}{with $\mathcal{L}_{S}$}                                                                                                        &            &            &            &               \\ \cmidrule(lr){1-2}
  \multicolumn{2}{c}{1st Encoder}  & 5.81     & 5.67      & 5.58     & 10.33            \\
  \multicolumn{2}{c}{2nd Encoder}  & 6.29     & 6.14     & 6.05     & 10.48           \\
  \multicolumn{2}{c}{3rd Encoder}  & 5.43     & 5.27     & 5.27     & 10.05            \\
  \multicolumn{2}{c}{4th Encoder}  & 4.37     & 4.55      & 4.35      & 8.75       \\ \\[-1em] \hline 
  \\[-1em]
  \\[-1em]
  \multicolumn{2}{c}{with $\mathcal{L}_{O}$}                                                                                                        &            &            &            &               \\ \cmidrule(lr){1-2}
  \multicolumn{2}{c}{1st Encoder}  & 5.08     & 4.88     & 5.09     & 9.00           \\
  \multicolumn{2}{c}{2nd Encoder}  & 7.10     & 7.31     & 7.48     & 10.51            \\
  \multicolumn{2}{c}{3rd Encoder}  & 4.64     & 4.36      & 5.00    & 9.27            \\
  \multicolumn{2}{c}{4th Encoder}  & 4.57     & 4.45      & 4.54     & 8.93  \\ \\[-1em] \hline 
  \\[-1em]
  \\[-1em]
  \multicolumn{1}{c|}{\ \ $\mathcal{L}_{S}$ \ \ } & $\mathcal{L}_{O}$    &       &        &        &      \\ \\[-1em] \cline{1-2} \\[-1em] 
  \multicolumn{1}{c|}{\multirow{4}{*}{\begin{tabular}[c]{@{}c@{}}4th\\ Enc.\end{tabular}}} & 1st  & \textbf{4.29}  & \textbf{4.11} & \textbf{4.15} & \textbf{8.67} \\  
  \multicolumn{1}{c|}{}                  & 2nd  & 4.60 & 4.43 & 4.38 & 9.09 \\  
  \multicolumn{1}{c|}{}                  & 3rd  & 4.48 & 4.19 & 4.35 & 8.81 \\ 
  \multicolumn{1}{c|}{}                  & 4th  & 4.47  & 4.2  & 4.41  & 8.81 \\
  \bottomrule[1.2pt]
  \end{tabular}
\end{table}

\section{Results and analysis}

\subsection{Performance comparison}
Table \ref{tab:der_comparison} shows the performance of SA-EEND \cite{fujita2019end2}, RX-EEND \cite{yu2022auxiliary}, and the proposed method in terms of DER.
For the SA-EEND \cite{fujita2019end2} and RX-EEND \cite{yu2022auxiliary} models, both the DERs reported in \cite{fujita2019end2,yu2022auxiliary} and those obtained from our own implementations are included in the table for comparison purposes.
To investigate the effect of applying the proposed auxiliary losses, either $\mathcal{L}_{S}$ or $\mathcal{L}_{O}$ was applied to one of the encoder blocks.
As there were two speakers in the training datasets, $\mathcal{L}_{S}$ was applied to two SA heads according to the criterion described in Section \ref{sec:proposed:training}, whereas $\mathcal{L}_{O}$ was applied to one SA head.
As shown in the table, the application of the SVAD loss ($\mathcal{L}_{S}$) to the 4th encoder block significantly improved the baseline system in both datasets, but the performance change was marginal for the other cases.
Unlike the SVAD loss, the OSD loss improved the baseline system across the different encoder blocks, except for the 2nd block that exhibited degraded performance for both datasets.
Interestingly, both SVAD and OSD losses showed worst performance when applied to the second encoder block.

Based on the results of the simulated datasets, we used both auxiliary losses in the SA-EEND model, as expressed in Eq.\,(\ref{eqn:L_total}), with $\alpha=1$ and $\beta=1$.
As shown in the last row of Table \ref{tab:der_comparison}, the application of both SVAD and OSD losses leads to a performance generally superior to that obtained from either one of the auxiliary losses.
The best performance in both datasets was obtained when the SVAD loss and OSD loss were applied to the fourth and first encoder blocks, respectively.
Our best model was also competitive with the recently proposed RX-EEND \cite{yu2022auxiliary} model for both simulated and real datasets.

\begin{figure}[t]
  \centering
  \includegraphics[width=\linewidth]{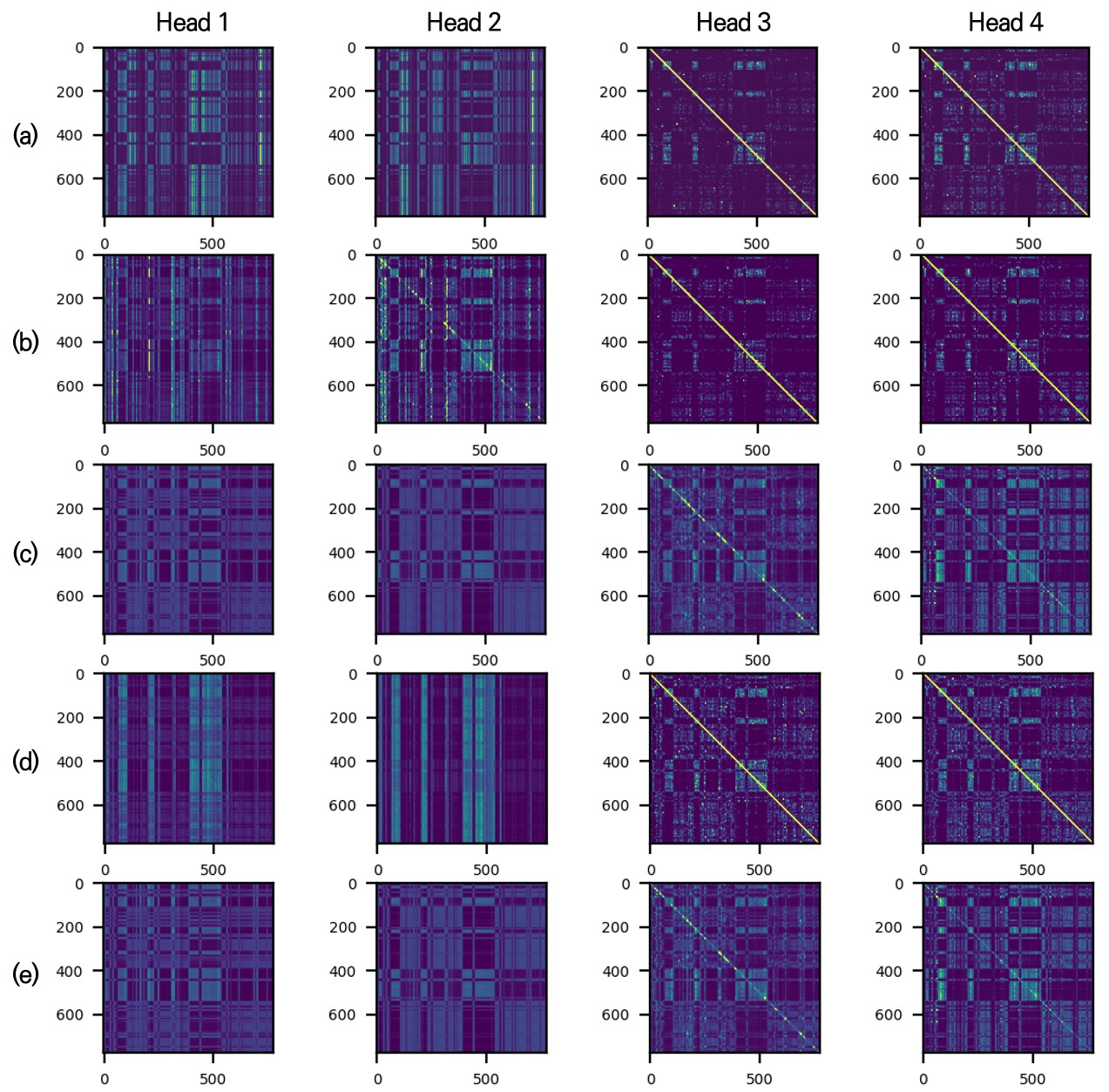}
  \caption{Visualization of attention weight matrices in the 4th encoder block: (a) SA-EEND, (b) RX-EEND, (c) SA-EEND with SVAD loss (d) SA-EEND with OSD loss, and (e) SA-EEND with both auxiliary losses applied to the 4th encoder block. (CH recording id: iaai)}
  \label{fig:head_visualized}
\end{figure}

\subsection{Visualization of SA heads}
\begin{figure}[]
  \centering
  \includegraphics[width=\linewidth]{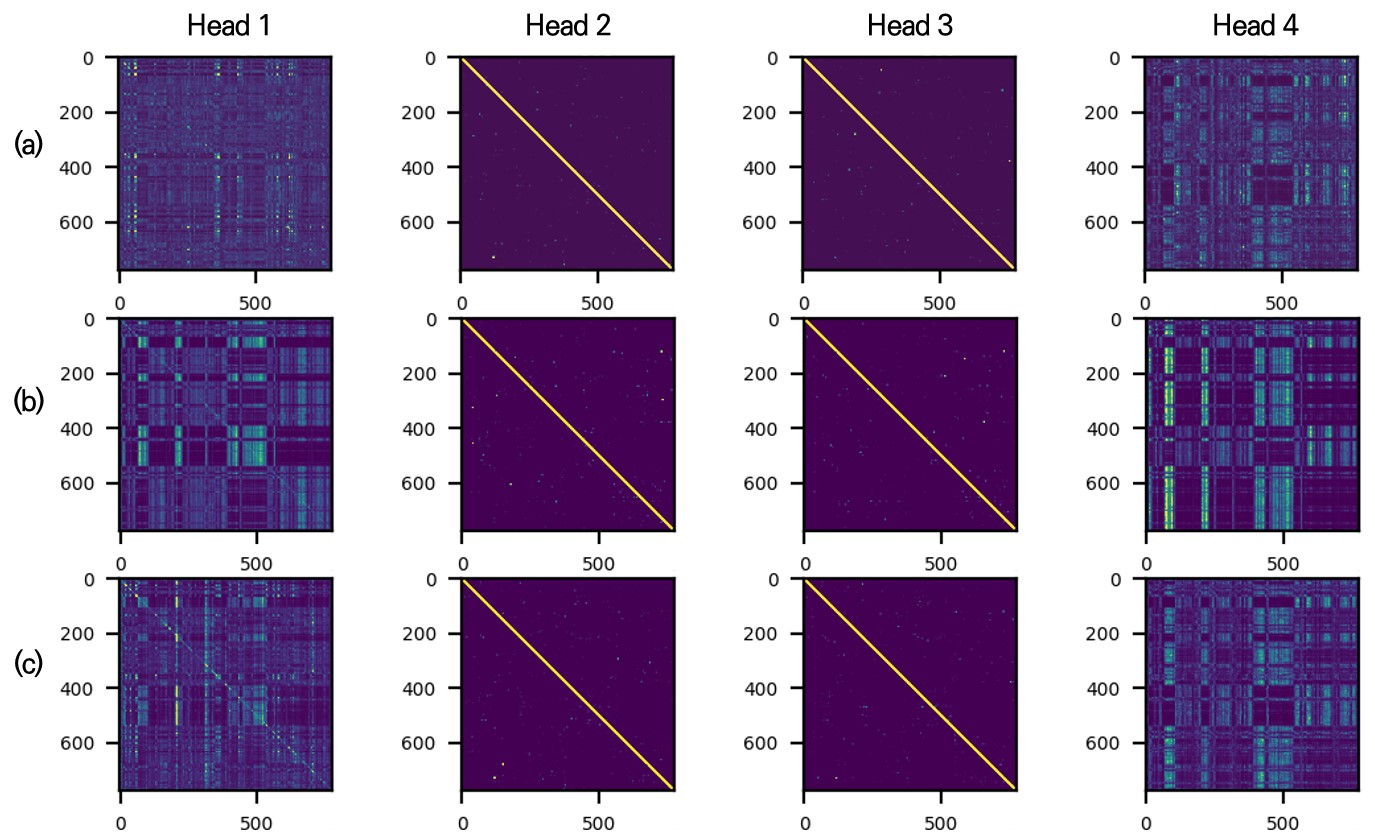}
  \caption{Visualization of attention weight matrices in the 2nd encoder block: (a) SA-EEND, (b) RX-EEND, (c) SA-EEND with both auxiliary losses applied to the 4th encoder block. (CH recording id: iaai)}
  \label{fig:SA_RX_Proposed}
\end{figure}

To investigate the effect of proposed auxiliary losses on the MHSA in the transformer encoder block of the SA-EEND model, the attention weight matrices (i.\,e., $A$ in Eq.\,(\ref{eqn:attention})) of the four SA heads are visualized in Fig.\,\ref{fig:head_visualized}.
In the figure, the attention weight matrices were calculated in the 4th encoder block, as they were located closest to the output layer that produces the final speaker posteriors.
Comparing the first row with the third, fourth, and fifth rows, the non-diagonal elements of the attention weight matrices corresponding to heads 3 and 4 showed clearer patterns with larger weights after the adoption of the proposed auxiliary losses.
Thus, considering the DER improvements presented in Table \ref{tab:der_comparison}, it can be inferred that the diversification of the identity-like patterns in the attention weight matrices achieved by applying the proposed auxiliary losses lead to more effective training of the SA-EEND model.
Interestingly, for heads 1 and 2, the patterns observed in Fig.\,\ref{fig:head_visualized}(c) and \ref{fig:head_visualized}(d) were different from those observed in Fig.\,\ref{fig:head_visualized}(a).
In Fig.\,\ref{fig:head_visualized}(e), which corresponds to the SA-EEND model subjected to both the proposed auxiliary losses applied to the fourth encoder block, the patterns of attention weight matrices were similar to those in Fig.\,\ref{fig:head_visualized}(c).
This suggests that when both the proposed OSD and SVAD losses were applied to the same encoder block, the SA-EEND model was more affected by the SVAD loss than the OSD loss.
Finally, it is interesting to see that for heads 3 and 4, the patterns obtained from the RX-EEND, as shown in Fig.\,\ref{fig:head_visualized}(b), were similar to those obtained from the baseline SA-EEND in Fig.\,\ref{fig:head_visualized}(a).
This could be because the performance improvement of the RX-EEND was mainly attributed to the auxiliary BCE losses applied to the lower encoder blocks in addition to the last block.
To examine this, the attention weight matrices calculated in the 2nd encoder block of the various EEND models are visualized in Fig.\,\ref{fig:SA_RX_Proposed}.
Notably, the patterns observed for the RX-EEND model (Fig.\,\ref{fig:SA_RX_Proposed}(b)) had larger weights compared to the others (Figs.\,\ref{fig:SA_RX_Proposed}(a) and \ref{fig:SA_RX_Proposed}(c)), while being somewhat similar to the patterns obtained from the 4th encoder block of the SA-EEND model (Fig.\,\ref{fig:head_visualized}(b)).
Interestingly, for all models presented in Fig.\,\ref{fig:SA_RX_Proposed}, the two SA heads always exhibited identity-like attention weight matrices, which assigns mostly equal weights for all frames.
This may explain the performance degradation cases of the proposed auxiliary losses, as presented in Table \ref{tab:der_comparison}, when either $\mathcal{L}_{S}$ or $\mathcal{L}_{O}$ was applied to the 2nd encoder block of the SA-EEND model.

\subsection{Ablation study on selection of SA heads}
As described in Sections \ref{sec:intro} and \ref{sec:proposed:training}, we assumed that 
the attention weight matrices that are similar to an identity matrix could be redundant and used this assumption to select the SA heads to apply the proposed auxiliary losses.
To validate the aforementioned head selection method, we trained the SA-EEND models using the proposed auxiliary losses but with random selection of SA heads for comparison.
For random head selection, we simply applied the proposed SVAD or OSD losses to the first or first two SA heads, respectively, without considering the similarity between the attention weight matrices and identity matrix.
Table \ref{tab:der_head_select} compares the results obtained from random head selection and proposed method described in Section \ref{sec:proposed:training} on the CH dataset.
The results indicated that the identity matrix-based selection of SA heads to which the proposed auxiliary losses are applied is effective, particularly when the SVAD loss was employed.

\renewcommand{\arraystretch}{0.92}
\begin{table}[t]
\caption{DERs (\%) on CALLHOME for different SA head selection methods. $\mathcal{L}_{s}$ and $\mathcal{L}_{o}$ were applied to the 4th encoder block.}
  \label{tab:der_head_select}
  \centering
\renewcommand{\tabcolsep}{5mm}
{\normalsize
\begin{tabular}{cccc}
\toprule[1.2pt]
\multicolumn{2}{c}{\textbf{Loss}} & \multicolumn{2}{c}{\textbf{Selection method}} \\ \cmidrule(lr){1-2} \cmidrule(lr){3-4}
$\mathcal{L}_{s}$     & $\mathcal{L}_{o}$     & Random matrix    & Identity matrix   \\ \\[-1em] \hline \\[-1em] \\[-1em]
\Checkmark        &             & 9.30      & \textbf{8.75}              \\
                  & \Checkmark  & 9.00      & \textbf{8.93}              \\
\Checkmark        & \Checkmark  & 9.22      & \textbf{8.81}              \\ 
\bottomrule[1.2pt]
\end{tabular}
}
\end{table}

\section{Conclusions}
In this study, we proposed training the SA-EEND model with auxiliary losses to better utilize the SA heads exhibiting patterns of attention weight matrices similar to an identity matrix.
The auxiliary losses were defined by exploiting the speech activity labels relevant to VAD or OSD, both of which can be considered important subtasks for speaker diarization.
To validate the idea, the proposed auxiliary losses were applied to the SA heads whose patterns of attention weight matrices were similar to an identity matrix.
The experimental results on the two-speaker diarization task showed that the proposed SVAD and OSD losses could significantly improve the performance of the conventional SA-EEND model in both simulated and real conditions.
The visualization and comparison of attention weight matrices for the different EEND variants supported our assumption and analysis.

\section{Acknowledgements}
This work was supported by Institute of Information {\&} communications Technology Planning {\&} Evaluation (IITP) grant funded by the Korea government(MSIT) (No.\,2021-0-00456, Development of Ultra-high Speech Quality Technology for Remote Multi-speaker Conference System)

\bibliographystyle{IEEEbib}
\bibliography{strings}

\end{document}